\def\lsim{\mathrel{\rlap{\lower4pt\hbox{\hskip1pt$\sim$}}
    \raise1pt\hbox{$<$}}}         
\def\gsim{\mathrel{\rlap{\lower4pt\hbox{\hskip1pt$\sim$}}
    \raise1pt\hbox{$>$}}}         
\pdfoutput=1

\newenvironment{myindentpar}[1]%
{\begin{list}{}%
         {\setlength{\leftmargin}{#1}}%
         \item[]%
}
{\end{list}}

\title{The Scientific Life of John Bahcall}

\author{Wick Haxton \\
Institute for Nuclear Theory and Department of Physics,\\
University of Washington, Seattle, WA 98195 USA}
\date{\today}
\documentclass[11pt]{article}
\usepackage{graphicx}
\usepackage{epsfig}
\usepackage{cite}
\usepackage[margin=2.5cm]{geometry}

\begin{document}
\maketitle

\begin{abstract}  
This article follows the scientific life of 
John Norris Bahcall, including his tenacious pursuit of the solar
neutrino problem, his contributions to our understanding of galaxies, 
quasars, and their emissions, and his leadership of and advocacy 
for astronomy and astrophysics.
\end{abstract}

\section{John's Best Day}
\label{sec:intro}

John Bahcall once mentioned that perhaps the best day of his scientific
career was one in 1964 when he received a call from Art Poskanzer,
a nuclear chemist then working at Brookhaven National Laboratory.  
The call came at the
urging of Ray Davis, who wanted John to know that Poskanzer's team --
as well as the team of Hardy and Verrall from McGill University -- had shown that the beta decay lifetime of $^{37}$Ca was indeed quite short.  The ensuing conversation earned Art's group a bottle of
champagne -- carried by Willy Fowler to Brookhaven a few months later, and opened in Director Maurice Goldhaber's office.

\begin{figure*}
\begin{center}
\includegraphics[width=13cm]{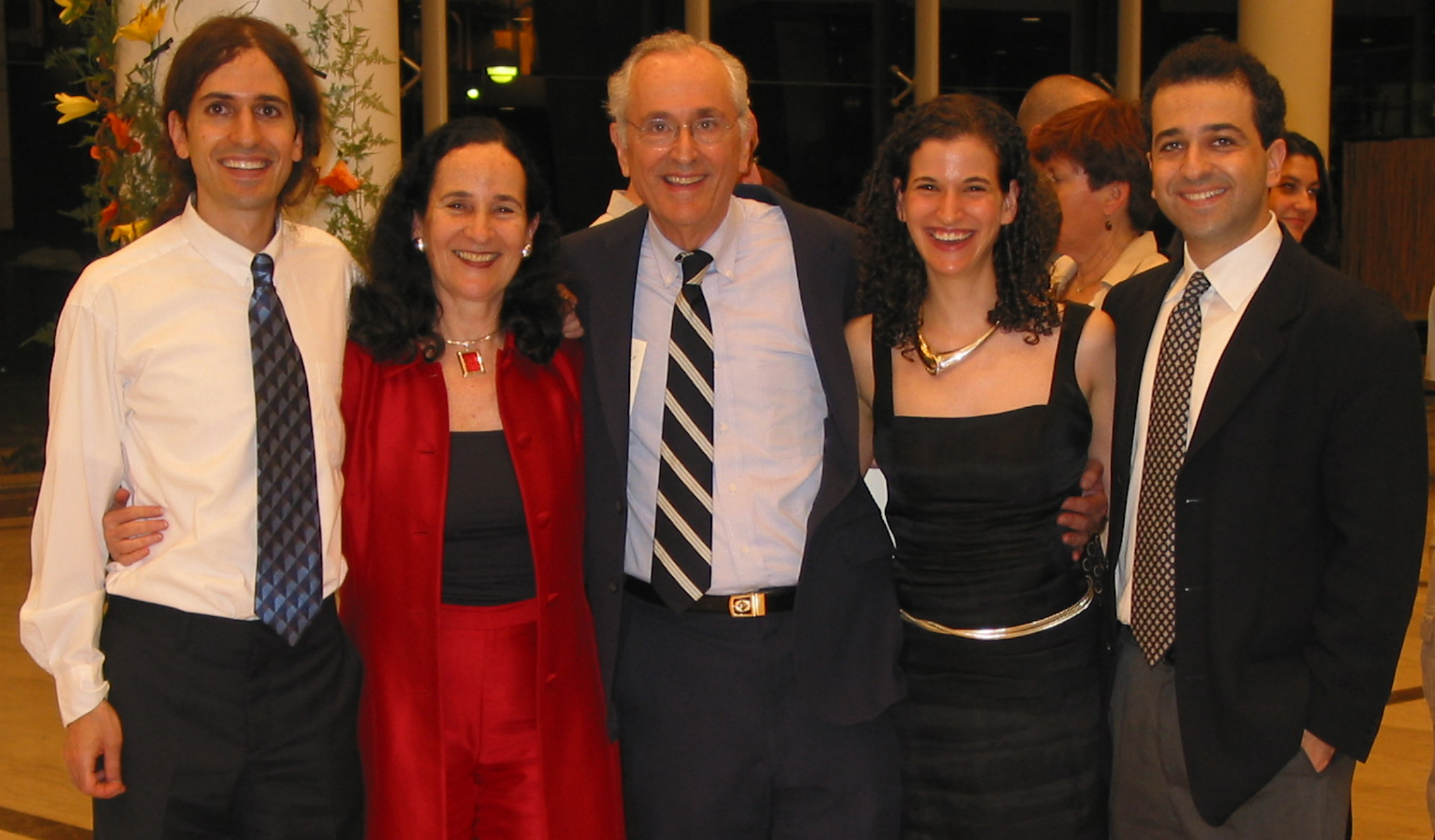}
\caption{John and  Neta Bahcall, with their children, Safi, Dan, and Orli, in Israel in 2003,
on the occasion of the presentation of the Dan David Award to John.  This was one
many times when John was honored by his colleagues: others include the
National Medal of Science (1998);  the Warner Prize of the American Astronomical
Society (1970); the NASA Distinguished Public Service Medal (1992); the Dannie 
Heineman Prize (1994); the Hans Bethe Prize (1998); the Russel Prize of the American
Astronomical Society (1999); the Benjamin Franklin Medal in Physics (2003);
the Gold Medal, Royal Astronomical Prize (2003); the Fermi Award (2003); and the Comstock
Prize of the National Academy of Sciences (2004).  ( Photograph courtesy of
the Institute for Advanced Study.)} 
\label{fig:DanDavid}
\end{center}
\end{figure*}

\begin{figure*}
\begin{center}
\includegraphics[width=13cm]{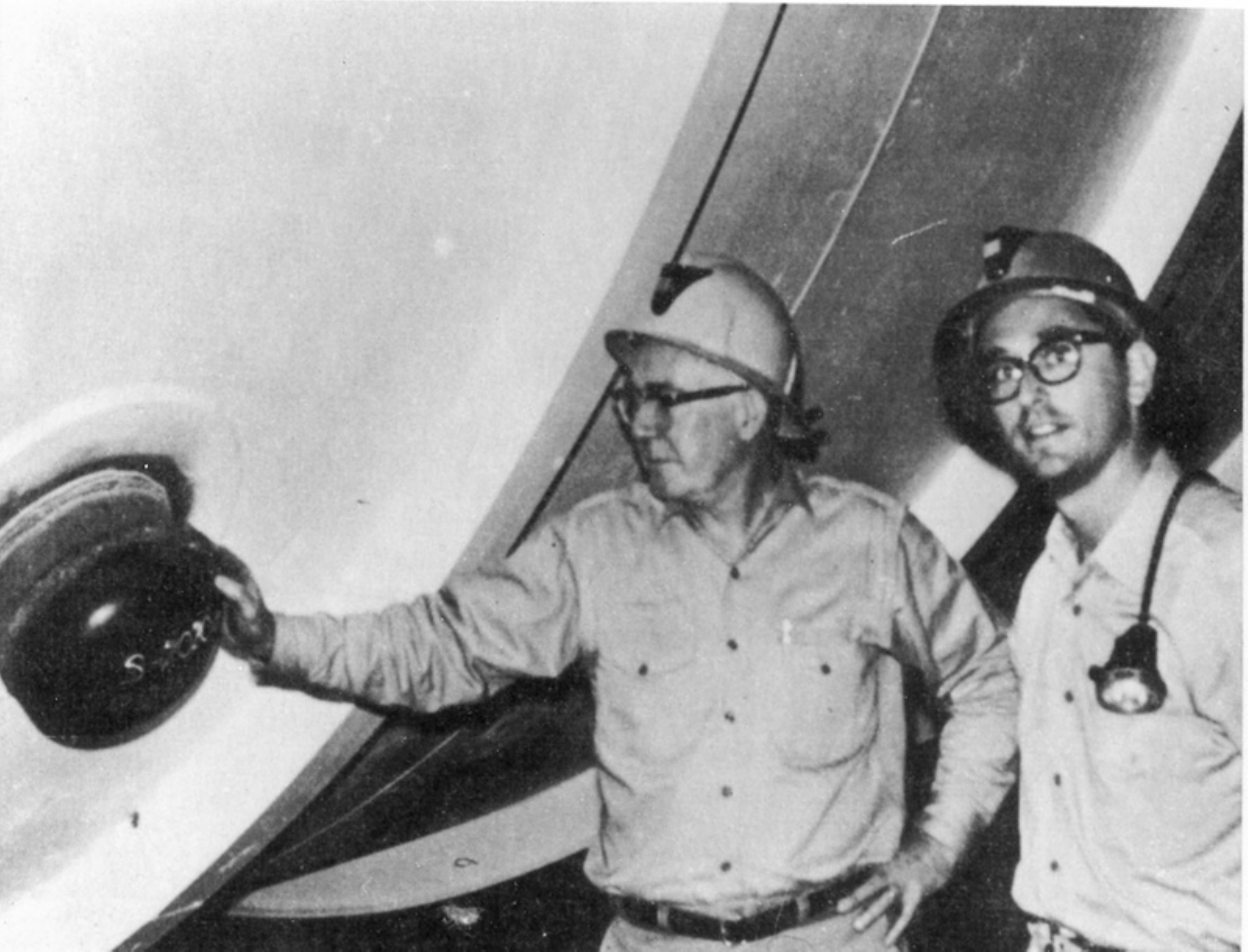}
\caption{Ray Davis and John Bahcall at the Homestake Mine, circa late 1960s.  Photograph courtesy of the Institute for Advanced Study. } 
\label{fig:JohnRay}
\end{center}
\end{figure*}

John Bahcall (Fig. \ref{fig:DanDavid}) was one of the dominant figures in 20th century 
astrophysics and the intellectual leader of the 40-year effort to 
understand the physics behind the solar neutrino problem.
Part of John's genius was his deep appreciation for the 
importance of teamwork in physics: he recognized that the goal
of understanding the Sun required experimentalists and theorists
to chip away at many obstacles, some astrophysical, others nuclear
and atomic.  An accomplished tennis player (Louisiana state champion),
John had learned early in life that one point
could make a game, one game could determine a set, and one set
could decide a match.  John knew that the lifetime of $^{37}$Ca
was a crucial point in a very important doubles match.  His
doubles partner was Raymond Davis Jr.,  and the trophy would be a
neutrino detector, deep within the Homestake Gold Mine in Lead,
South Dakota (Fig \ref{fig:JohnRay}).

John's role in the Homestake effort and in many solar neutrino
endeavors that followed was that of both player and coach.  His
personal research drove the effort to accurately model the Sun,
to understand its seismology and neutrino fluxes, and to exploit the
neutrino flux as a test for new physics.  But he also advocated
for and helped focus many other activities that were important to
the quest to solve the solar neutrino problem. 
This advocacy was crucial to the new generation of detectors that
ultimately led to the discovery of neutrino oscillations.

John's interests and influence ranged far beyond solar physics.
He authored nearly 500 technical papers, making major contributions
to galactic modeling and structure, quasars, and the
production of ultra-high-energy neutrinos.  He wrote or edited
nine books, including ``Neutrino Astrophysics" \cite{Book} and ``The Red Shift
Controversy" \cite{Arp}.  He was
one of the most recognized spokespeople for astronomy and astrophysics,
frequently interacting with the media, and publishing approximately
sixty popular science articles.   He led one of the world's
premier astrophysics programs at the Institute for Advanced Study, Princeton. 
There he mentored generations of young postdocs and fellows -- nearly 
300 of the field's finest young researchers -- recruiting them,  following their
progress in research with close attention, and helping many
find good positions when the time came to leave the IAS.  He presided over
the Tuesday astronomy luncheon (now the Bahcall Lunch),
and its traditional grilling of visitors and locals.  Together with
Lyman Spitzer, Jr., he was a tireless advocate for
the Hubble Space Telescope and the
Space Telescope Science Institute. He chaired the ad hoc committee that
jumpstarted the effort to create DUSEL, the Deep Underground 
Science and Engineering Laboratory to be sited at Homestake.
He served as President of the
American Astronomical Society, led the 1990 National
Research Council's decadal survey for research and instrumentation in
astronomy (the ``Bahcall Report"), and was President-elect of the
American Physical Society.

\section{Solar Neutrinos}
\subsection{The Early Days \cite{Account}}
\begin{figure*}
\begin{center}
\includegraphics[width=13cm]{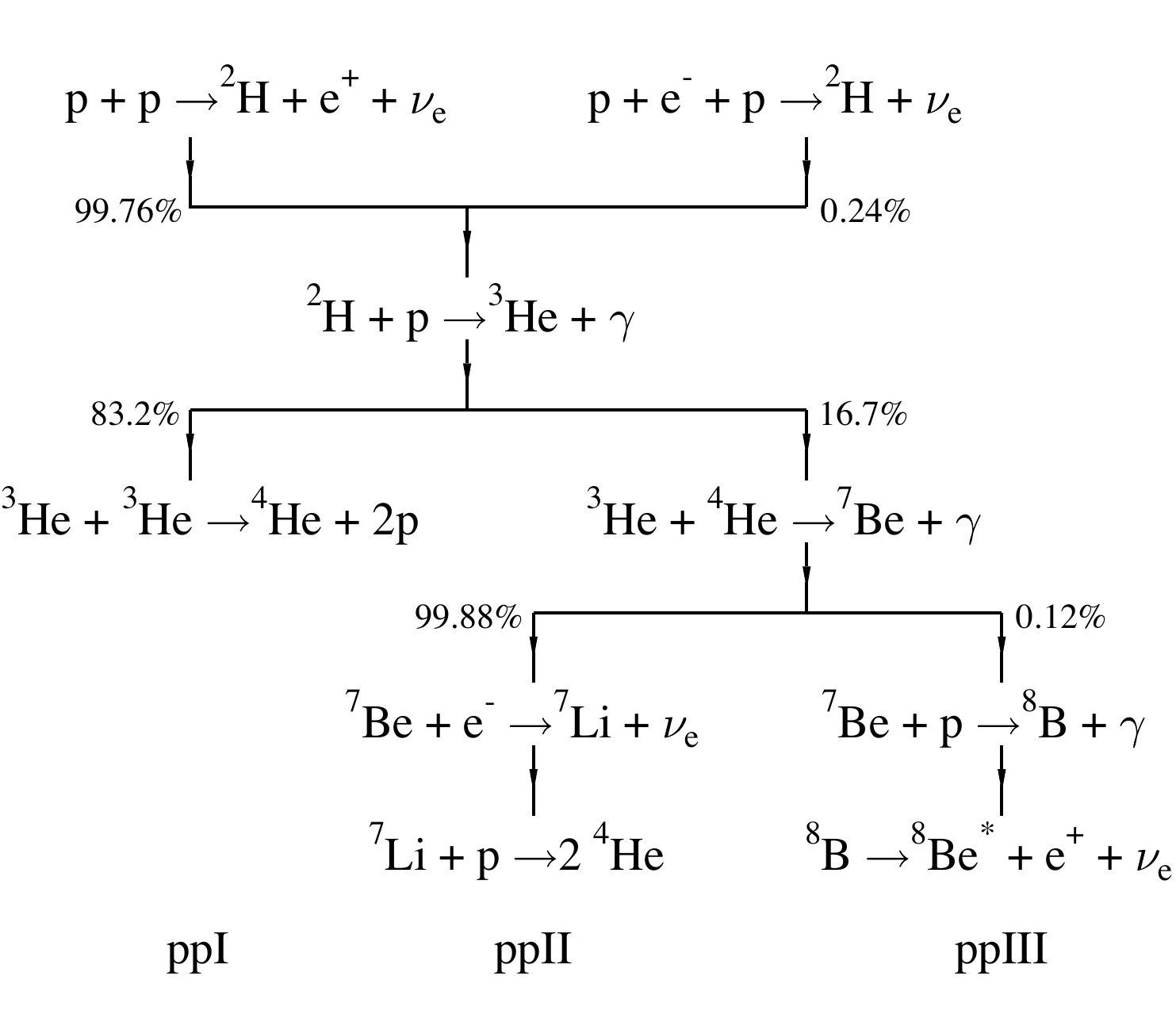}
\caption{The pp chain.  Each of the three cycles, ppI, ppII, and ppIII, is associated
with a characteristic neutrino source.  The competition between the cycles depends
sensitively on the solar core temperature.} 
\label{fig:pp}
\end{center}
\end{figure*}

In 1958 Holmgren and Johnston \cite {HJ58,HJ59} found that the cross section for
$^3$He + $^4$He $\rightarrow$ $^7$Be + $\gamma$ was about 1000 times
larger than expected, implying that the solar pp chain for synthesizing
$^4$He would have additional terminations beyond the ppI end reaction 
$^3$He + $^3$He $\rightarrow$ $^4$He + 2p (see Fig. \ref{fig:pp}).  
Higher energy neutrinos from the new cycles fed by $^3$He + $^4$He, 
the ppII and ppIII cycles of Fig. \ref{fig:pp}, could be measured by the techniques Davis had developed at Brookhaven \cite{Davis55}.
This was discussed by A. G. W. Cameron \cite{Cameron} and 
by W. A. Fowler \cite{WAF58}.  Fowler noted that quantitative predictions
required more information on reactions involving $^7$Be,
namely the rates for p + $^7$Be, 
then unmeasured, and e$^-$ + $^7$Be.   

At the time John Bahcall was a graduate student at Harvard, having
finished undergraduate studies at UC Berkeley in 1956 and his masters
degree at the University of Chicago in 1957.  In 1961 John began 
a postdoctoral appointment -- a year with Emil Konopinski 
at Indiana University, working on weak interactions in nuclei. 
There John wrote a paper describing the temperature
and density dependence of beta decay in stellar interiors.  
Willy Fowler, who refereed the paper for the Physical Review, described the 
results to Davis, who then wrote to Bahcall in February 1962 to inquire 
about the rate for electron capture on $^7$Be in the Sun.  Bahcall
and Davis note, in their entertaining ``An Account of the Development
of the Solar Neutrino Problem," that this question was the first of many
they were to ask of each other in the decades to come \cite{Account}.  Bahcall's paper on bound-state and 
continuum electron capture on $^7$Be appeared in the Physical
Review later that year \cite{JNB62}.

Davis's inquiry reflected his interest in measuring
neutrinos from the Sun,
using a detection scheme suggested by Pontecorvo \cite{Pontecorvo} and further 
explored by Alvarez \cite{Alvarez}, $^{37}$Cl$(\nu,e)^{37}$Ar.  In 1955 Davis completed
an exploratory experiment at Brookhaven -- 1000 gallons of C$_2$Cl$_4$ buried 19 ft under ground -- which set an upper limit on the rate of solar neutrino captures
of $\sim$ 40,000 SNU [solar neutrino unit=
$10^{-36}$ captures/target atom/s, a term John coined].  If the pp chain were to operate
entirely through the ppIII cycle, Davis estimated that the
rate would be about 7.7 captures per
day, or 3900 SNU, easily within reach in an improved
experiment.  Thus it became important to understand the 
solar fate of $^7$Be, which depends both on nuclear physics and 
the Sun's central temperature.

The critical missing reaction rate, $^7$Be(p,$\gamma$)$^8$B, was measured
by Ralph Kavanagh in 1960 and found to be low \cite{Kavanagh60}.  It became clear
that the ppIII cycle was not dominant, and that the measurement of
the associated flux would be a daunting task.  Precisely how
daunting depended in part on the core temperature of the Sun.

In summer 1962 John became a research fellow in Willy Fowler's group 
at Caltech's Kellogg Laboratory, joining fellows Icko Iben and Dick Sears.
This group undertook the task of incorporating the new
information on the pp chain into a model of the Sun, in order to
estimate the flux of solar neutrinos for the first time.  As described by John \cite{Account}, 
Sears performed the calculations using an energy-generation subroutine
and opacity code originally developed by Iben, with the former adjusted
by Bahcall and Fowler to reflect several new results on cross sections
as well as John's formulation of
electron capture.  The model results were then used by John in
a hand calculation of the neutrino flux.  The model predicted an average
temperature for the central core of 1.5 $\times 10^7$ K and fluxes
for $^8$B and
861 keV $^7$Be neutrinos of 3.6 $\times 10^7$ and 1.0 $\times 10^{10}$
/cm$^2$s, respectively. The paper noted the extreme sensitivity of the
$^8$B flux to temperature \cite{BFIS}.

These fluxes, which later proved somewhat optimistic, were not encouraging.
Although Davis was contemplating a very large experiment, 100,000 gallons
of C$_2$Cl$_4$ housed in a cavity deep underground, the
expected counting rate was about an event per day.  That
estimate was based on a cross section for neutrino capture on $^{37}$Cl
derived from the known electron capture rate of $^{37}$Ar.  This rate,
the only experimental information available, corresponded to a transition
between the ground states of the parent and daughter nuclei.

During a seminar presented by John at the Niels Bohr Institute in summer 
1963, Ben Mottelson raised a question about the possible role of excited 
states in $^{37}$Ar.  John addressed this question by performing a calculation 
in which the states of $^{37}$Ar and $^{37}$Cl -- as well as the analog
nuclei $^{37}$K and $^{37}$Ca -- were modeled as $1d_{3/2}^3$ hole configurations
in a $^{40}$Ca core,
a treatment that generates in $^{37}$Ar a T=3/2 excited state, the 
isospin analog of the $^{37}$Cl ground state, and thus a superallowed 
transition to this state \cite{JNBcross}.  Consequently he found a total cross
section for $^8$B neutrinos nearly 18 times that of the ground-state
cross section: the $^{37}$Cl experiment would be primarily 
sensitive to neutrinos from the highly temperature dependent ppIII cycle.  Most
important, apart from small corrections due to charge-symmetry violation,
the same physics would govern the analog beta decay $^{37}$Ca $\rightarrow$
$^{37}$K, greatly shorting the half-life to a predicted 0.13 s.  While
$^{37}$Ca had not yet been observed, it was apparent that there was
an experimental method for confirming the calculation, and thus
demonstrating the potential for the chlorine detector to measure $^8$B
neutrinos.  John estimated that his calculation would likely be 
reliable to $\pm$ 25\%.

In March 1964 Bahcall \cite{JNBPRL64} and Davis \cite{DavisPRL64} published companion articles in Physical Review
Letters arguing that the envisioned 100,000-gallon experiment, if 
conducted deep underground, would succeed in measuring solar neutrinos and
thus in determining the temperature of the solar core.  Bahcall discussed
uncertainties in detector cross sections and solar model flux
predictions, concluding that the expected counting rate would be 
(40 $\pm$ 20) SNU.  Davis reported the results of a 1000-gallon pilot
experiment that had been conducted at a depth of 1800 m.w.e. [meters of water equivalent] in an
Ohio limestone mine, and argued that a 100,000-gallon experiment 
conducted at 4000 m.w.e. would record between 4 and 11 
solar neutrino events per day, with backgrounds a factor of
10 of more below this level.

John learned from Poskanzer's call that $^{37}$Ca had been discovered.
Its measured half-life, 0.17 s, was short and within the 25\% uncertainty John had estimated
for his calculation \cite{Hardy64,Reeder64}.  John recognized
immediately that this confirmation of one of the key assumptions in the Bahcall
and Davis companion papers placed the chlorine experiment on firm
ground.  It would be possible to probe the Sun's core with solar neutrinos.

\subsection{The Standard Solar Model}
In a follow-up paper to the Bahcall, Fowler, Iben, and Sears neutrino flux calculation,
Sears explored the sensitivity of the
predictions -- notably $\phi(^8$B) -- to input
assumptions, such as the solar composition \cite{Sears64}.  Sears 
postulated a homogeneous zero-age-main-sequence Sun
(because the proto-Sun likely passed
through a fully convective Hayashi phase), then fixed the initial
heavy element abundance Z to the observed (but then poorly
determined) surface abundances.
He adjusted the He/H ratio, Y/X, to reproduce the Sun's 
luminosity after 4.5 b.y.  He found that Z and Y were
correlated, with the lowest Z explored (0.020) yielding the
lowest primordial helium abundance (Y=0.272) and lowest $\phi(^8$B),
1.9 $\times 10^7$/cm$^2$s.  The standard solar model (SSM) 
neutrino flux sensitivity to input parameters would concern
John and his colleagues for four decades.

In April 1968 Davis, Harmer, and Hoffman announced an upper bound
on the solar neutrino capture rate of $\lsim$ 3 SNU, so that $\phi(^8$B) $\lsim$
0.2 $\times 10^7$/cm$^2$/s \cite{DDH68}, based  on the
initial two runs of the Cl detector.
The baseline for comparison to theory
came from the SSM calculation of Bahcall and Shaviv \cite{BS68}, which
was scheduled for publication in the Astrophysical Journal,
having gone through final revisions in January 1968.
This paper reflected 
important progress in the nuclear physics of the pp chain,
including Parker's remeasurement \cite{Parker66} of the critical $^7$Be(p,$\gamma$)$^8$B reaction and new measurements that substantially increased
the cross section for $^3$He + $^3$He $\rightarrow$ $^4$He + 2p
\cite{Nengming66, Bacher67, Winkler67}.
It also identified the heavy element abundance and the rate for
p+p $\rightarrow$ D + $\nu_e$ + e$^+$ as key uncertainties affecting
the $^8$B flux prediction, $\phi(^8$B) $\sim 1.3(1 \pm 0.6) \times
10^{7}$/cm$^2$s. 

The comparison 
with experiment also depended on the cross section for neutrino absorption on $^{37}$Cl, which by this time was firmly established.
A suggestion by Charlie Barnes had led to a more precise relationship
between $^{37}$Ca beta decay and neutrino capture on $^{37}$Cl:
In 1964 Bahcall and Barnes pointed out that the model-dependent Gamow
Teller distribution for neutrino capture could be extracted from
the delayed proton spectrum following $^{37}$Ca($\beta^+$)$^{37}$K
\cite{Barnes64}.  By 1966 the
first measurement of the delayed protons had been made \cite{Poskanzer66}. 

Davis, Harmer, and Hoffman observed that their upper bound
was about a factor of seven below the SSM prediction.  Their letter
was accompanied by one from Bahcall, Bahcall, and Shaviv updating
the Bahcall and Shaviv results to reflect a new determination of Z
and the neutron lifetime (and thus the weak axial coupling $g_A$).  These changes lowered
the counting rate of their most probable model to 7.7 $\pm$ 3.0 SNU,
a range still outside the experimental limit \cite{BBS68}.

The experiment had given an unexpected answer,
and the wait for independent confirmation would take two decades.    The
result obtained by Davis, a limit of a fraction of a count per day in a massive volume of
organic liquid, seemed incredible to many.  Transcripts of discussions that occurred
during a 1972 conference on solar neutrinos in San Clemente \cite{SC72} show Davis responding
patiently to various skeptics, describing tracer and other cross-checks he had performed 
to verify the efficiency of the nearly single-atom counting.  

Similarly, on the theoretical side, a few years of effort on pp-chain nuclear physics and 
metalicities had reduced the SSM prediction from 40 to $\sim$ 8 SNU: was a remaining
factor-of-three discrepancy a serious matter?   As the necessary adjustment in the SSM
was a reduction in the Sun's core temperature by $\sim$ 5\%, ``no" was
perhaps not an unreasonable answer.   One early suggestion was
the ``low Z" model of Iben \cite{Iben69}, in which the convective zone's metals were attributed to
the accumulation of dust and other debris during main-sequence evolution, while the
core's much lower Z reflected the true composition of the primordial gas cloud.   Indeed,
Bahcall, Bahcall, and Shaviv had concluded ``if the usual theory of stellar interiors
is correct, then the heavy element abundance Z must be less than 2\% by mass in order
for the predicted neutrino-capture rate not to exceed the observed value."  Another
suggestion was the mixed Sun of Ezer and Cameron \cite{Ezer68}, which
kept the Sun's central opacity low by replenishing the core's hydrogen. 
Over the next 20 years many nonstandard solar models were proposed, with most designed
to solve the solar neutrino problem by producing a cooler Sun.

John's views about the SSM and the solar neutrino puzzle slowly evolved.
In his 1989 book ``Neutrino Astrophysics" \cite{Book}-- some 27 years after his initial work
on this problem -- he offered an explanation for the solar neutrino problem:
\begin{myindentpar}{0.5cm}
{\it Simplicity.  Our models of the solar interior and of neutrino propagation 
are not strongly constrained by experimental data.  My guess is that a decade of new
experiments will show that we need more sophisticated theoretical models,
astrophysical and physical.}
\end{myindentpar}
But John would remark, just a few years later, that perhaps the 
inclusion in his book of a chapter on nonstandard solar models had been
a mistake, a throwback to earlier times.  What changed his views?

John's exploration of
possible uncertainties in the SSM began in earnest with his
1969 paper with Neta Bahcall and Roger Ulrich \cite{BBU69}.  There was a great deal to explore: by one count the
modern SSM has 19 adjustable parameters, incuding the solar age and luminosity, 
individual metal abundances, nuclear cross sections, and the coefficient for diffusion.
The collaboration with Roger Ulrich
extended over eight papers and included studies of the consequences of changes in composition,
magnetic fields, and radiative opacity (a collaboration with the Los Alamos opacity group)
on SSM predictions.  It included detector absorption cross sections, as new ideas
emerged for followups to the chlorine experiment, and John's first paper on helioseismology
(which, interestingly, made the point that p-mode frequencies were then not a restrictive 
constraint on neutrino fluxes).   

(The collaboration with Neta, of course, transcended their 30 joint papers, encompassing
a lifetime of shared experiences and three children, Safi, Dan, and Orli, who gave
John much joy.  John,  describing the start of this lifelong collaboration \cite{abb}, 
told of meeting a graduate student on a trip to Israel in 1965
``with a beautiful smile that stole my heart." )

At the 1984 Homestake Solar Neutrino conference John expressed the view that
no good solution existed \cite{HSJNB}:
\begin{myindentpar}{0.5cm}
{\it The standard solar model predicts -- if nothing else happens to the neutrinos on the way
to the Earth -- about 6 SNU, with an effective 3$\sigma$ uncertainty of about 2 SNU.
This is in conflict with observations reported by Keith Rowley at this conference,
which yield about 2 SNU (with a small 1$\sigma$ uncertainty of about 0.3 SNU).  
There is no accepted solution for this problem, although many have been proposed...
The discrepancy between theory and observation has remained approximately
constant over the past 16 years, although there have been hundreds of careful and important
papers refining the input data, the calculations, and the observations.}
\end{myindentpar}
Roger Ulrich once noted \cite{HSRKU} the tendency ``for workers in each of the three areas related
to the [solar neutrino] problem -- stellar interior theorists, particle physicists, and
experimental physicists -- to hope and occasionally believe that the solution lay in the
other fellow's camp."    However, John's comment above shows a broader skepticism
about solutions, including those from particle physics.   In 1980 \cite{JNB3}
he expressed doubt about neutrino mixing scenarios because of the requirement of nearly
maximal mixing:
 ``...it is difficult to resolve the difference between predictions based on the
solar models and observations solely by invoking neutrino oscillations, if there are
only three kinds of neutrinos coupled to each other." 

Thus John's Homestake talk focused on the need for more measurements:
he presented his well-crafted argument that the proposed gallium experiment
was needed to separate the innocent from the guilty.
Because this experiment would be primarily sensitive to pp neutrinos, a
minimum astronomical counting rate of 78 SNU was guaranteed for any standard or 
nonstandard solar model
(assuming a steady-state Sun).  In contrast, neutrino oscillations would yield a counting rate
of about 38 SNU, since complete mixing of three neutrino flavors was needed to 
account for the Cl results.

The gallium experiment,
proposed by the Russian theorist V. A. Kuzmin \cite{Kuzmin}, was 
similar in approach  to chlorine, but involved more complicated chemistry.  Two
extraction procedures had been designed, one
for a GaCl$_3$-HCl solution and one for metallic liquid gallium.
Davis played a central role in both efforts.  A pilot experiment with 4.6 tons of GaCl$_3$
solution had been conducted at Brookhaven.  Although Bahcall passionately advocated for
a full-scale experiment, and although two high-level review committees (headed respectively
by Seaborg and Vandenbosch) recommended proceeding, a US experiment was
never mounted.   The cost of gallium procurement was the primary obstacle.
BNL's collaborators, after nearly eight years, moved ahead on
GALLEX as a primarily European effort, led by Till Kirsten \cite{GALLEX}.  The group
quickly obtained
commitments from three national science ministries.  Similarly, the 60-ton
Soviet-American Gallium Experiment (SAGE) was mounted at the Baksan
Laboratory, under
the direction of Vladimir Gavrin, George Zatsepin, and Tom Bowles \cite{SAGE}.  SAGE and
GALLEX began operations in January 1990 and May 1991, respectively.

While the gallium saga played out, efforts were underway to reduce backgrounds in
the Kamiokande proton decay detector, so that it could operate at a threshold below
the $^8$B neutrino endpoint.   The experimenters reached thresholds of 9 MeV and later
7.5 MeV.  Data from the first 450 days of running was reported in the July 1989 Physical
Review Letters \cite{KamiokandeA} -- a rate 46\% that expected for SSM fluxes: confirmation of the Homestake
results had taken two decades.
Kamiokande II/III operated from December 1985 until July 1993, accumulating
2079 live detector days of data \cite{KamiokandeB}.

\subsection{Neutrino Oscillations}
These experiments influenced John's views on the solar neutrino puzzle, but
so did new developments in theory.
In 1986 Mikheyev and Smirnov \cite{MS1985,MS1986} evaluated the effects of
matter on solar neutrino oscillation probabilities, using a result for the
effective neutrino mass originally derived by Wolfenstein \cite{Wolfa,Wolfb}.  The result, nearly
complete flavor oscillation of $\nu_e \rightarrow \nu_\mu$ even for quite small
mixing angles $\theta_{12}$, is known as the MSW effect.  The effect was rather
quickly recognized to be a consequence of adiabatic level crossing \cite{Bethe,Haxton,Parke}.  
Within the Sun charge-current scattering of a $\nu_e$ off electrons generates
a density-dependent contribution to the neutrino effective mass.  Consequently
solar neutrinos can encounter a critical density, on their way out of the Sun, where
this effective mass just cancels the vacuum mass difference between two neutrino
mass eigenstates.  The crossing of this critical density can generate a nearly
complete change in flavor, and thus low counting rates in detectors sensitive
only to (Cl, Ga) or primarily to (Kamioka II/III) electron neutrinos.

Thus an elegant particle-physics solution was available to account for the solar
neutrino problem, one that was viable even if neutrino mixing angles, like those
among the quarks, were small.

John joined Ray Davis, Jr. and Lincoln Wolfenstein in the 1988 Nature review ``Solar Neutrinos: A Field
in Transition," which summarized a solar neutrino conference that the
Institute for Theoretical Physics, Santa Barbara, had hosted the previous year \cite{KITP}.  The paper
discussed the MSW mechanism as well as very early results from Kamioka II
(which had already established an upper bound on the $^8$B neutrino flux in conflict with the SSM). 
The paper did not indicate which type of solution -- solar or particle physics --
the authors preferred.  Instead, it stressed that the field was undergoing a transition, driven by
new experiments.   In 1989, anticipating results from SAGE and GALLEX, John 
and the author explored MSW solutions to the solar neutrino
problem as a function of possible gallium outcomes, finding for a wide range of
counting rates, 20-100 SNU, that distinct ``islands" of solutions appeared in the 
$\delta m^2-\sin^2{2 \theta}$ plane -- adiabatic, nonadiabatic,
and large-mixing-angle solutions \cite{JNBWCH}.

The following year John and Hans Bethe decided to place their bets: in a Physical Review
Letter that was widely read, they argued that the solution was the MSW mechanism,
and chose from among possible MSW solutions nonadiabatic conversion with a small mixing angle \cite{Johnandhans}.   
Their first conclusion was correct, but the second step proved premature, as it would take
the field another decade to sort out the competing solutions.   While they were aware
that large-mixing-angle solutions would be 
allowed for gallium outcomes above 20 SNU, John and Hans argued that neutrino mixing
angles would likely be small, similar to those known from the quark mass matrix, thus
disfavoring this solution.
They also argued against a hybrid small-angle solution -- one where the
high-energy portion of the $^8$B flux experiences an adiabatic crossing, suppressing
these neutrinos, while the
lowest energy neutrinos reside in the nonadiabatic portion of the MSW triangle, and thus
would survive to be counted in SAGE and GALLEX -- because 
the Kamioka II rate for high-energy neutrino events was nearly half the SSM prediction.
John and Hans
pointed out that one consequence of their selected solution could be a very low 
gallium counting rate, perhaps as low as 5 SNU.

But the gallium experiments produced a different result.   The first runs from SAGE
\cite{SAGE1} yielded
a counting rate of 20$^{+15}_{-20} \mathrm{(stat)} \pm 32 \mathrm{(sys)}$ and a 90\% c.l.
upper bound of 79 SNU.  This result was compatible with any of the 
three oscillation solutions.  The first result from GALLEX \cite{GALLEX1} was a counting rate $83 \pm 19 \pm 8$ SNU ($1\sigma$),
leading the collaboration to note ``astrophysical reasons remain as a possible explanation
of the solar neutrino problem."  Indeed, the final SAGE and GALLEX results were to converge
to values quite close to John's minimum astronomical rate, 78 SNU.  

In 1993, with the gallium results in hand, John and Hans
asked the question \cite{Imply} ``Do Solar Neutrino Experiments Imply New Physics?"
They answered that the general pattern
of fluxes derived from the Cl, Kamioka II, and GALLEX/SAGE experiments ``suggest[s] that
new physics is required beyond the standard model" but cautioned that all of their
arguments ``depend to some extent on our understanding of the solar interior."  They
relied on extensive Monte Carlo studies of SSM uncertainties in reaching these
conclusions.  John began a new series of SSM investigations, anchored by his
collaboration with Marc Pinsonneault, but also including David Guenther, Sarbani Basu, 
J\o rgen Christensen-Dalsgaard,  Aldo Serenelli, and others.   The work included
the effects of helium and heavy-element diffusion on the SSM:  such corrections became
important as the quality of the data from helioseismology improved.  This allowed the
modelers to test SSM predictions against precisely known data, including the depth
of the convective zone and the Sun's interior sound speed.   The SSM sound
speed was found to agree with helioseismology to better that 0.2\% throughout almost the entire
Sun.  John argued in several contexts that this was perhaps a more severe test of the 
SSM than solar neutrino spectroscopy.  He was enormously excited about this SSM
success, concluding that it was now very likely that the solution to the solar neutrino
problem would be new physics.

\begin{figure*}
\begin{center}
\includegraphics[width=13cm]{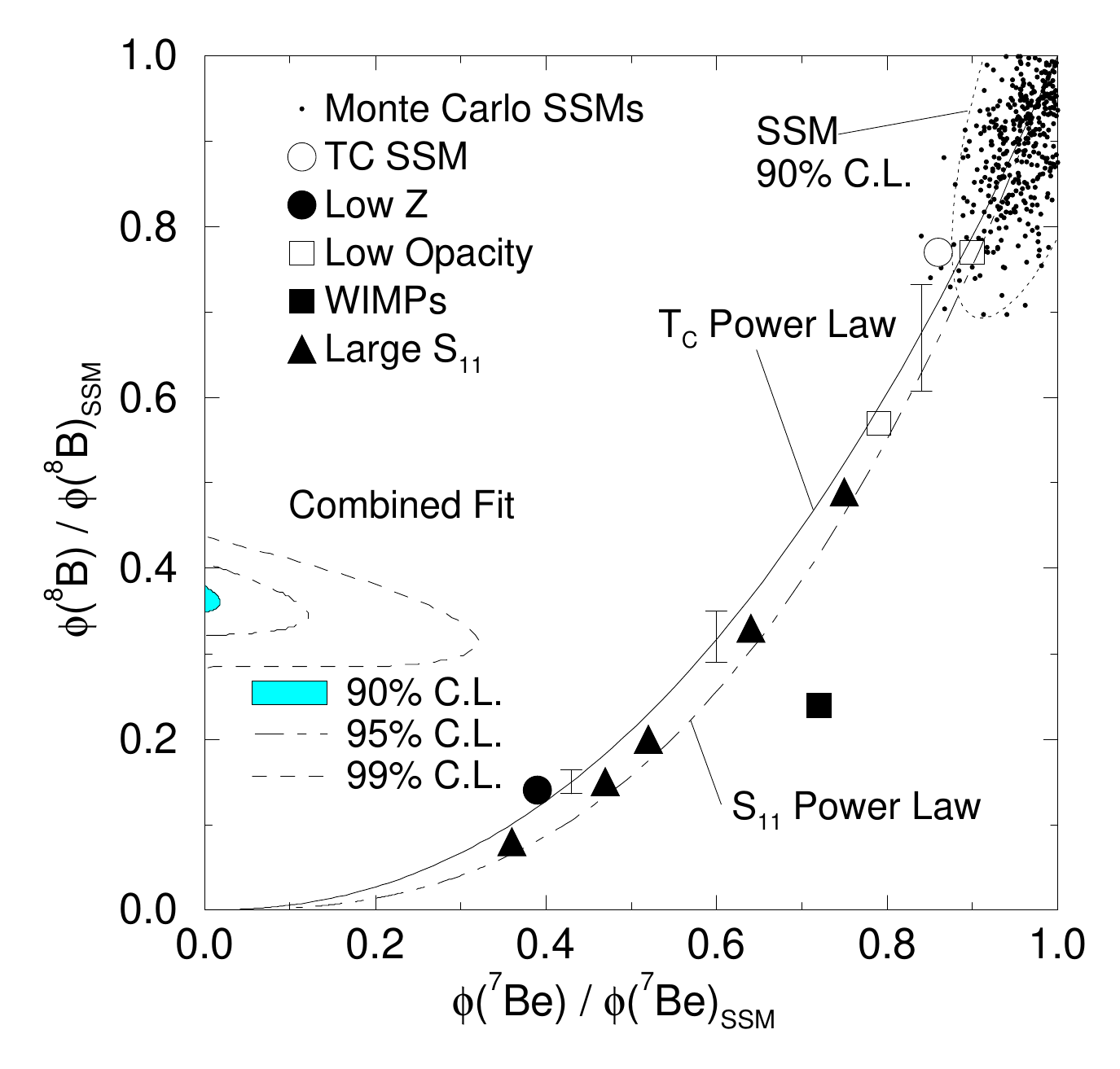}
\caption{A figure from \cite{Hata} illustrating the inconsistency between measured
neutrino fluxes and the general trends in standard and nonstandard solar models,
a pattern of neutrino fluxes reflecting the solar core temperature.} 
\label{fig:hata}
\end{center}
\end{figure*}

Further evidence that the SSM might not be at fault came from
the neutrino flux systematics revealed by the experiments.  While 
many solar model ``dials" can be turned, in the
end the predicted neutrino fluxes are strongly correlated with
one solar property, the core temperature.  If one lowers
the temperature by a few percent, $\phi(\mathrm{pp})$ is nearly unchanged (provided
the model is constrained to reproduce the luminosity);  $\phi(^7\mathrm{Be})$ is
somewhat suppressed; and $\phi(^8\mathrm{B})$ is significantly suppressed.  But
the results from Kamioka II/III and GALLEX/SAGE  seemed to require
the $^7$Be neutrino flux  to be the most sharply suppressed.  A nice illustration of the
conflict between the measured fluxes and trends based on the solar core temperature
is given in Fig. \ref{fig:hata}, from Bludman, Hata, and Langacker \cite{Hata}.

On June 5, 1998, the Super-Kamiokande collaboration announced \cite{SK98} evidence for 
neutrino mass, attributing an anomaly seen in atmospheric neutrino data to oscillations.
While this discovery was not directly related to solar physics -- the results were consistent with
$\nu_\mu \leftrightarrow \nu_\tau$ oscillations and corresponded to a $\delta m_{23}^2$
too large to generate an MSW crossing in the Sun -- it was clearly a game changer.
The neutrino sector included the basic ingredients needed for a particle-physics solution 
of the solar neutrino problem, massive neutrinos and flavor mixing.  

On June 30, 2001,
the Sudbury Neutrino Observatory (SNO) collaboration
submitted a paper on its initial charged-current (CC) and
elastic scattering (ES) results to
Physical Review Letters \cite{SNO1}: assuming no flavor mixing, the two rates were found
to be inconsistent at 3.3$\sigma$.   A second letter, submitted on April 19, 2002, 
provided dramatic evidence for oscillations:  the solar  neutrino flux was measured
independent of flavor, through the neutral-current  (NC) breakup of deuterium \cite{SNO2}.  The three channels,
CC, ES, and NC, agreed and pointed to a solution in which two-thirds of the
$^8$B neutrinos arrive at Earth as heavy-flavor neutrinos.  The SNO collaboration deduced
a total flux, $(5.09^{+0.44}_{-0.43} \mathrm{(stat)}^{+0.45}_{-0.43} \mathrm{(sys)} )\times
10^8$/cm$^2$s, consistent with the SSM prediction.   Ironically, the mixing angle
$\theta_{12} \sim 35^\circ$ is large -- Nature chose not
to enhance the effects of a small mixing angle through the MSW mechanism.

Forty years earlier Bahcall, Iben, and Sears had taken on a formidable task, construction
of a model of the Sun that could quantitatively predict neutrino fluxes.  John stayed with
this task, recognizing its significance.  The model's input parameters were refined 
through many years of patient measurement by nuclear and atomic experimentalists:
John's advocacy helped keep these communities motivated.
Improvements in the quantum mechanics governing opacities was needed, and new
effects like diffusion had to be incorporated in the model, in response to
the new data from helioseismology.   When the model converged, John assessed its
precision in careful studies, and became convinced of its accuracy.   Finally, when the
SNO collaboration finished its work, he had the pleasure of an answer, one that rewarded 
his hard work and demonstrated its lasting importance.

In a television interview \cite{NOVA} John was asked about how he reacted to the results
from SNO.  ``I was called right after the
[SNO] announcement was made by someone from the {\it New York Times} and asked
how I felt.  Without thinking I said `I feel like dancing I'm so happy.' ... It was like a person
who had been sentenced for some heinous crime, and then a DNA test is made and
it's found that he isn't guilty.  That's exactly the way I felt." 

\section{Beyond the Sun}
The solar neutrino problem was John's life-long scientific passion, but it certainly did
not define the boundaries of his interests.  John made lasting contributions to a wide range of
problems at the forefront of physics and astrophysics, including
low-energy weak interactions, quasar absorption line analyses, galactic modeling, 
dark matter, neutron star structure and cooling, the stellar environments around massive black holes,
and high-energy cosmic neutrinos.   Indeed, such problems comprise the majority of
his scientific work.

\subsection{Beta Decay and Electron Capture}  
John once noted that Konopinski's ``Theory of Beta Radioactivity" was perhaps the
book that had influenced him most.  In 1961, as a postdoc working in Emil's group, 
John sat in on a course based on the notes that became this book, making
up his own practice problems, and solving and publishing them.  His understanding
of weak interactions in atoms and nuclei provided a foundation for his life's work.

John's involvement with the solar neutrino problem
began with his paper on beta decay in stellar interiors.  In the first few years of his career
he explored several aspects of weak interactions, often not in connection with
astrophysics.   Topics included bound-state beta decay,
relativistic effects in atoms, electron capture rates from higher atomic states, and exchange
and overlap effects in beta decay.   The last topic deserves special note, as
John's paper \cite{JNBoverlap} contains several results that have survived the test of time.

Twenty-five years after this paper was published,
the possible discovery of a 17 keV neutrino caused some excitement.  The initial experiment revealed
an anomaly in the electron spectrum from tritium beta decay which, because of the 18.6 keV 
energy release,
appeared within 2 keV of the endpoint.   Thus the outgoing electron is very slow: this is
a limit where exchange corrections -- the decay electron goes into a bound atomic state,
while the spectator atomic electron is shaken off -- become large.  The author used John's
1963 exchange corrections to estimate the effects, which accounted for part of the observed anomaly.
On seeing the paper, John called to express his pleasure that his ancient paper had not
been forgotten.

Test of quark unitary depend on determinations of $V_{ud}$ from $0^+ \rightarrow 0^+$
$\beta$ decay.  These experiments require extraordinary precision, including very accurate
determinations of the mass difference between the parent and daughter atoms.   For example,
Penning trap experiments have determined some mass differences to 50-100 eV.
But the masses measured are those of the stable atoms. 
High-energy-release $\beta$ decays
are essentially instantaneous, so that the daughter nucleus inherits the atomic configuration
of the parent state.  The energy loss in the subsequent atomic arrangement, a correction that
should be applied to the Penning trap results, can exceed 100 eV in heavy atoms.  
This correction has recently been made in Fermi $\beta$ decay analyses \cite{Hardy},
using John's formula.

\subsection{QSO Absorption lines}
Quasars -- quasi-stellar radio sources -- were identified as very distant sources by
Maarten Schmidt and collaborators in 1962, when one such object (3C273) was shown to reside at a
redshift of 0.158 \cite{Schmidt,Oke}.  Their emissions include a continuum that extends into 
far ultraviolet and X-ray frequencies.  Quasars, or more generally, quasi-stellar objects (QSOs),
were the focus of much of John's work in the late
1960s and then again three decades later, when the Hubble Space Telescope opened
up new opportunities.

By 1967 QSOs were at the forefront of astronomy, recognized, because of
Schmidt's work, to be significantly more common in the early universe than now.   (The QSO
distribution peaks at redshifts of two to three.)
As powerful, early sources, they were potentially an important probe of cosmology and 
cosmological changes.   In particular, the discovery of the first quasar with 
redshift greater than two, 3C9, opened up some spectacular possibilities \cite{Schmidt2}.  Gunn and
Peterson \cite{Gunn} showed that, for such a distant source, an absorption trough could
appear in the continuous spectrum, if there was sufficient neutral hydrogen in the
intergalactic medium.  This feature would be apparent in spectra measured from the Earth's
surface, if red-shifted into optical wavelengths.  The trough is formed through absorption at the
wavelength of Lyman-$\alpha$ photons, as the light travels over cosmological distances.

Very soon after the Gunn and Peterson paper,
Bahcall and Ed Salpeter wrote an Astrophysical Letter ``On the Interaction of Radiation
from Distant Sources with the Intervening Medium" in which they envisioned a scenario
where the intervening gas was clumpy \cite{Lines}.    The trough would then be replaced
by a series of sharp absorption lines, displaced from the Lyman-$\alpha$ wavelength
by the redshifts of the gas clumps.  That is, the intergalactic medium could be probed at
a variety of distances by the pattern imprinted on the spectrum of a distant QSO.  The following
year the authors published a second letter describing how, under favorable conditions,
the wavelength, depth, and width of the absorption lines could constrain the temperature,
chemical composition, and velocity dispersion of cluster gas \cite{Lines2}.  This letter noted that
Bahcall, Peterson, and Schmidt had begun an examination of Schmidt's QSO spectra for
this purpose.  A few months later, the group described hydrogen and carbon absorption 
lines found in the spectra for a quasar at redshift 2.118, corresponding to absorption
at redshift 1.949 \cite{Lines3}. 

In 1969 John teamed with Lyman Spitzer on the letter ``Absorption Lines Produced by
Galactic Halos," proposing, for the first time, that many of the QSO absorption lines
with multiple redshifts are caused by gas in large extended halos around normal
galaxies \cite{JohnSpitzer}.  This suggestion came long before the existence of
such halos had been established by other observational means. 

John worked intensively on the QSO absorption line problem for five years, producing about
30 publications.   The early data were not of high quality, and there were questions
about distinguishing absorption that might be associated with the QSO's immediate 
environment from that connected with intervening gas clouds.   John attacked this 
problem by carefully examining the data and by numerical modeling, assessing in Monte
Carlo studies the significance of multiple redshift systems, each of which might
be characterized by multiply detected absorption lines.  His was the first
quantitative approach to absorption line analysis -- replacing previous ``by eye"
identifications with an algorithm-based analysis that accounted for line strengths, 
expected line ratios, and other systematics.  The times were exciting, as QSOs were
new and appeared to open a doorway to the very distant universe.  Part of that excitement
was captured in a famous debate between John and
Chip Arp on whether quasar redshifts were
cosmological \cite{Arp} -- a debate John clearly won.

John's studies included
correlations in QSO directions with those of galaxy clusters,
bounds on the masses of QSOs,  and the question of whether the fine structure constant 
might vary with cosmic time --
a topic of significant interest today.  (The possibility that $\alpha$ might not be constant
had been suggested by Gamow, as a factor that could complicate conclusions drawn
from redshifts about steady-state vs. expanding universes.)   This issue was re-examined
in follow-up papers \cite{Alpha1,Alpha2} on the constancy of several metal line splittings for
QSOs and radio galaxies, ultimately yielding $\alpha(-2 \mathrm{b.y.})/\alpha(\mathrm{today})=
1.001 \pm 0.002$.  John's last paper on this subject was published in 2004 \cite{Alpha3}.

John's later advocacy for the Hubble Space Telescope reflected in part his early
interests in QSOs: high quality spectra could be obtained in the UV above the atmosphere,
with very low background, thereby extending both the range of distances that could be 
probed and the number of lines that might be correlated for a given source.  In
particular, the HST was crucial in probing QSOs in the low-redshift universe.   When the HST was carried
into space in 1990, John returned to absorption line astronomy with a passion, as
Principal Investigator of the HST Quasar Absorption Line Survey.   The Survey collaborators studied
some 80 lines of sight, effectively probing matter along each direction through redshift,
making use of the Faint Object Spectrograph.   The catalogs of Lyman-$\alpha$ and
metal absorption lines provide a detailed map of structure from the very nearby 
universe out to z $\gsim$ 3.  John was able to return to many of the themes he had explored
in the late 1960s -- such as the correlations of absorbers with galaxies -- but with much 
finer data and within the context of contemporary 
efforts to model the formation and evolution of structure.

In the end, John's work spanned four decades and deeply influenced our understanding
of QSOs and cosmological structure.   He was involved in the key questions --  the
identification of QSOs with cosmological distances, their correlation with galaxies and clusters
of galaxies, the utility of quasar light in absorption as a probe of the distribution of neutral hydrogen
and metals, QSO number density evolution with redshift,  QSO masses, 
and the role of QSOs as the central engines of host galaxies.

\subsection{The Bahcall-Soneira Model of the Galaxy}
In 1980 John Bahcall and Ray Soneira \cite{BSon1,BSon2} constructed a phenomenological model
of the Milky Way with parameters that were adjusted to reproduce stars counts made
in various directions in the sky, and which then would provide a basis for extrapolating
those counts (e.g., to fainter populations).   This model was extensively developed by
Bahcall and Soneira in the early 1980s, and has remained in broad use ever since.
It was motivated in part by the anticipated  Hubble Space Telescope program 
to probe the universe at unexplored faint magnitudes.   A reliable model 
of the galactic environment could help this program in many ways, such as in
assessing, from stellar trends in the Milky Way, how unknown local stellar populations
might interfere with cosmological surveys.

The model combined a thin exponential
disk with a spheroid, or bulge,
associating with each a distinct stellar population.   The spheroidal component was assumed
to be dominated by Population  II stars (typically older, less luminous, cooler, and relatively
metal-poor), similar to those found in globular clusters.
The empirical model for the disk assumed a broad range of stellar populations ranging
from extreme Population I (hotter, younger, relatively metal-rich) to Population II.
The stellar luminosity functions and density scale heights were assumed to be fixed 
throughout the galaxy -- that is, they were not allowed to vary with distance from the
galactic center.   The functional forms for star density were derived from observed
light distributions in external galaxies: the parameters included the scale height and length 
describing the exponential fall-off of the star density perpendicular to or along the galactic plane.
Specific values for these parameters were determined from galactic observations,
with the scale height allowed to vary with absolute magnitude, following
observational trends (e.g., a larger scale height for disk dwarfs, and a smaller one
for disk giant stars).  Similarly,  the distribution of stars as a 
function of luminosity and their color-magnitude relations were
determined for each of the two populations by fitting to local observations. 

The model was then tuned by exploring small variations in the initial parameters, to
identify the values that could provide the best star counts and colors in various observational
directions.  This corrected for correlations among parameters that might not
have been properly reflected in the initial values.

John and Ray made use of two observational data sets that were rapidly growing in the 1980s:
the global large-scale properties of external galaxies -- e.g., trends
in luminosity as a function of distance from the galactic center or height above the disk
midplane -- and the local properties of our region of the Milky Way, including its
luminosity density and metalicity.   Their model provided a way to encode a great deal of observational
data into a relatively small number of parameters, and then to test the adequacy of the
parametrization with further data. 

\begin{figure*}
\begin{center}
\includegraphics[width=13cm]{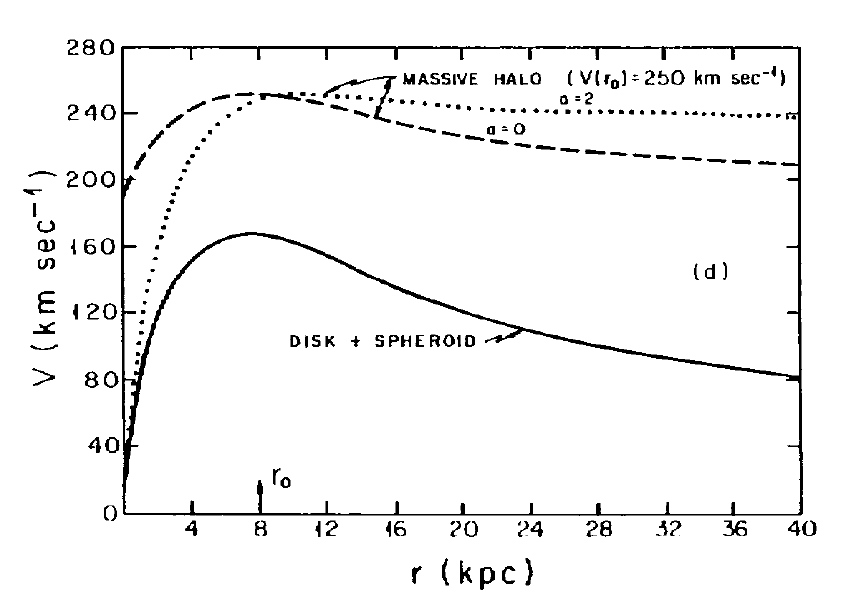}
\caption{A figure from \cite{BSon2} showing the Milky Way's rotation curve for
the Bahcall-Soneira model, contrasted with the flatter rotation curve that could
be achieved under the assumption of an additional and otherwise ``dark" massive halo component.} 
\label{fig:BSoneira}
\end{center}
\end{figure*}

The model was used to test the consistency of star counts with predicted galactic properties..
One of the first calculations done by John and Ray concerned the rotation
curve predicted for the Milky Way: the disk+spheroid model fell monotonically beyond 12 kpc, while
external galaxies of the same morphology have flat or slightly increasing rotational curves 
out to $\gsim$ 30 kpc \cite{BSon2}.  They showed that a massive halo component could be added 
to produce both the correct local rotational properties
and a flat rotation curve out to distances in excess of 40 kpc (see Fig. \ref{fig:BSoneira}).  They further argued
that if this population were stellar, it would have to be a new class of very faint stars that
did not follow star-count trends of the known stars.  
The surveys later done by the HST showed that no such halo population of faint dwarf
stars was present in the Milky Way.  In work with Hut and Tremaine \cite{BHT}, John used the
model's characterization of wide binaries to constrain any unobserved low-mass
stellar population in the disk.  (Oort had argued \cite{Oort} that the amount
of mass in the disk might exceed that observed in disk stars.)
By considering the disruption of wide binaries by the passage of low-mass stars, they
showed that any significant population of such stars was ruled unless the typical
stellar mass was less than 2M$_\odot$.

The Bahcall-Soneira model served as a standard model for the Milky Way for two decades, and
remains in use today.  It was arguably the first highly successful descriptive model of the
Milky Way.  The model formed a bridge between early observations and today's
more detailed numerical models 
of the Milky Way and other galaxies.

\subsection{Compact Objects and Stars: The Bahcall-Wolf Cusp}
Richard Wolf, as a graduate student at Caltech, wrote a term paper for Willy Fowler's
nuclear astrophysics course.  Willy advised Dick to talk with John (then a postdoc,
but soon to be an assistant professor at Caltech) about the paper and other topics.
Dick became John's first student, and the term paper led to Dick's first publication,
on the termination of the proton-proton chain at high densities \cite{Wolf}.
Dick completed his thesis in 1965 on the rates of
nuclear reactions in white-dwarf stars and on neutron star cooling.  
Over a twelve-year period he and John collaborated on eight additional papers, including the
Bahcall-Wolf model for stellar clusters in the vicinity of a black hole \cite{BWolf}.

Their work on neutron stars, occurring at the same time Bahcall and Davis were building
their arguments for the chlorine experiment, was motivated by the possibility that
neutron stars could be important galactic X-ray sources.  The work focused on the early cooling
of a hot neutron star by neutrino emission, including both the modified Urca process
of Chiu and Salpeter \cite{Chiu}
\[ n + n \rightarrow n + p + e^- + \bar{\nu}_e \]
and the effects of a possible pion condensate
\[ \pi^- + n \rightarrow n + e^- + \bar{\nu}_e. \]
Inclusion of the latter process -- which they discovered could greatly increase the cooling
rate of hot neutron stars -- was remarkable, given that work on pion condensation
in neutron stars did not begin until the early 1970s \cite{Migdal,SS}.
Sawyer and Scalapino note, in their first paper, that
``In almost all theoretical treatments the matter has been assumed 
to consist entirely of fermions, that is, baryons and leptons," but modify this statement with
a footnote to Bahcall and Wolf: ``However, see J. Bahcall and R. A. Wolf ... for a suggestion
that pions may enter and for some of the consequences for neutron stars."

But the two best-known papers by Bahcall and Wolf were on another topic, the 
distribution of stars around a massive black hole.   This question
was raised by Wyller \cite{Wyller} and then investigated semi-analytically by Peebles
\cite{Peebles1,Peebles2}.  In their first paper Bahcall and Wolf derived an evolution equation
for the diffusion of stars in the $1/r$ field of a black hole under certain simplifying
assumptions: 1) a stellar distribution that is spherically symmetric in coordinate space,
approximately isotropic in velocity space, and described by a one-body distribution;
2) equal mass stars; 3) the stellar mass is small compared to the black hole mass, 
which is small compared to the globular cluster mass; and 4) a star is destroyed by star-star collisions
or when its binding energy in the potential well exceeds a specified critical value.  The
resulting diffusion equation was solved numerically, showing that the stellar distribution
function evolves rapidly to an equilibrium configuration on a timescale determined
by the mean stellar collision time. 
The resulting equilibrium stellar density distribution around the
black hole was found to vary as $r^{-7/4}$ \cite{Cusp1}.  They argued that this characteristic distribution might
allow one to identify black holes within nearby globular clusters, provided the mass
of the black hole $\gsim 5 \times 10^3$ M$_\odot$, or $\gsim 10^3$ M$_\odot$
given a large space telescope.  In their second paper they generalized the treatment to
an arbitrary spectrum of stellar masses, finding similar results \cite{Cusp2}.  In the case of a system of 
stars with two different mass they found a shift in the power law index
from $\gamma$ = 7/4 to $\gamma$ = $m_1/4 m_2 + 3/2$, with $m_1<m_2$, suggesting that
the sharpness of the cusp might vary from 3/2 to 7/4, depending on the breadth of the
stellar mass distribution.

The possibility that a cusp in the stellar distribution might be a signature of a
central black hole remains a topic of great interest.   In recent years direct N-body simulations
for stellar systems containing a massive central object have verified
the time-dependence (configuration space and phase space) predicted by the Fokker-Planck
equation and produced a Bahcall-Wolf density cusp with $\rho \propto r^{-7/4}$ \cite{Preto}.
Observers have exploited the opportunity provided by our own galaxy to study the structure
and dynamics of stars in the vicinity of Sgr A*, the massive black hole at the center of the
Milky Way.  They have found some trends in agreement with the Bahcall-Wolf predictions,
but also  cusps with less severe slopes  \cite{Schodel} -- presumably because
one of the assumptions of the Bahcall-Wolf analysis is not satisfied by Sgr A*.

\subsection{High Energy Neutrinos: The Waxman-Bahcall Bound}
In the late 1990s John became interested in a variety of questions concerning high energy
astrophysical neutrinos, including their sources, propagation, and connection with 
hadronic cosmic rays.  One reason for his interest was the prospect that such
neutrinos would soon be seen in a new generation of massive detectors.  For example
IceCube, a high energy neutrino telescope nearing completion in the Antarctic, uses
strings of phototubes to view a cubic kilometer of deep, clear ice.   Neutrinos may be
the ultimate tool for probing the high-energy limits of the universe:  they 
propagate over cosmological distances unimpeded by fields or matter.

In 1997 John began work in this area, collaborating with
Eli Waxman, who was completing a five-year research stay at the Institute
for Advanced Study.  Among the papers produced
by Eli and John were two concerned with defining the high energy neutrino fluxes that 
might plausibly be produced in various astrophysical explosions.   The upper limit they
placed on such fluxes is known as the Waxman-Bahcall bound \cite{Wax1,Wax2}

\begin{figure*}
\begin{center}
\includegraphics[width=13cm]{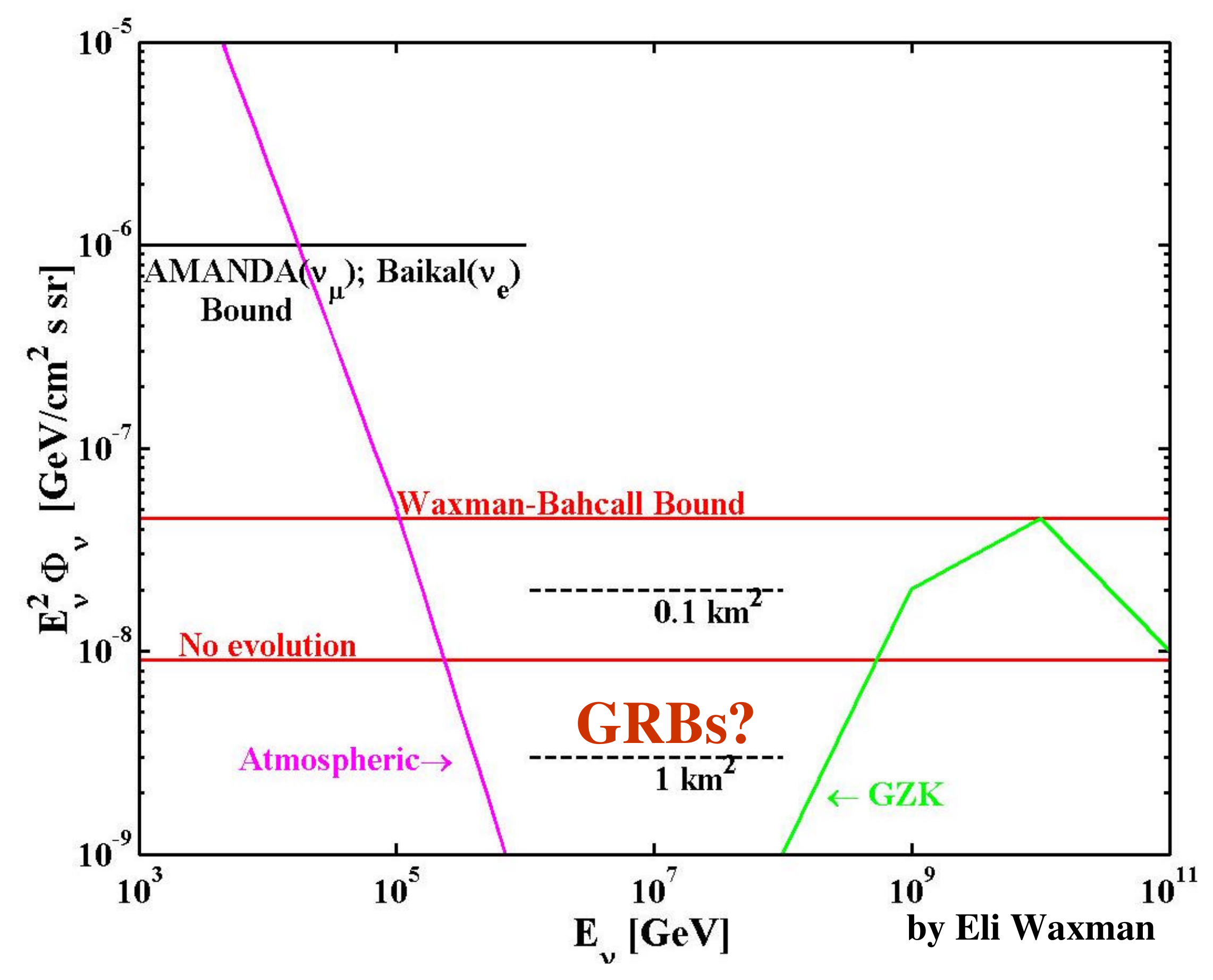}
\caption{A viewgraph from John Bahcall's September 2000 talk at the International
Conference on Neutrinos and Subterranean Science showing the implications of
the Waxman-Bahcall bound for various neutrino telescopes.}
\label{fig:waxman}
\end{center}
\end{figure*}

The candidate astrophysical sources of very high energy neutrinos include objects like
active galactic nuclei (AGNs) and the fireballs responsible for gamma ray bursts.  
The Waxman-Bahcall bound was based on arguments that the sources of high energy
neutrinos would also produce high energy protons.  The fluxes 
of hadronic cosmic rays are known for energies $\lsim$ 10$^{20}$ eV, the point at
which their propagation is restricted by the cosmic microwave background.   AGNs are
likely sources of energetic protons because their powerful, extensive jets
accelerate charged particles.  These protons then produce high-energy neutrinos: 
sources include the decays of pions and kaons produced by photoproduction off protons,
as well as proton-proton bremsstrahlung.  

The Waxman-Bahcall bound applies to sources
that are optically thin for high energy protons with respect to
meson-nucleon and photo-meson interactions.   With this assumption,
one can relate the resulting neutrino flux produced by such sources to the 
corresponding high-energy cosmic ray proton flux, since energetic protons can also 
leave the source.  This is the basic idea behind the Waxman-Bahcall bound.  
From the observed flux of high-energy cosmic ray protons at Earth and the calculated
fraction of energy lost by protons to pions, Eli and John derived the bound
\[ E_\nu^2 \phi_\nu \lsim 2 \times 10^{-8} { \mathrm{GeV} \over \mathrm{cm}^2 \mathrm{s~sr}}. \]

At the time the bound was published, more optimistic estimates of neutrino fluxes from
AGNs and other sources were in the literature.  Thus this bound generated some
discussion, a situation John generally relished.  In terms of detector sensitivity, the bound
converts to a minimum detector mass of about 0.1 gigatons -- if the bound is saturated.
Thus neutrino telescopes such as IceCube (1 gton) may be sufficient to detect the very
high energy neutrino flux \cite{IceCube} (see Fig. \ref{fig:waxman}).

\section{Reflections and Acknowledgements}
This attempted summary of John's scientific life is woefully inadequate.  It is a
one-dimensional projection of a career with many more dimensions.   Those who knew John 
recognize his impact goes far beyond the few selected topics described here.  He was
devoted to his work, with endless enthusiasm and energy for the problems he felt 
needed to be solved, and with a vision that often extended decades beyond the present.
But beyond this personal involvement in science, he had a very special gift
for helping focus others on the important work,
and of making his scientific goals a community destiny.   Those of us who worked on the 
solar neutrino problem always knew what we were doing was important -- because
it was important to John.   I do not know how to express this other than to say -- it
does not feel the same, now that he is gone.

On Saturday, October 29, 2005, the Institute for Advanced Study hosted a Tribute to 
John Bahcall.   I owe a great debt to the speakers who contributed to the Tribute's scientific session,
Art McDonald, Carlos Pe\~ na-Garay, Eli Waxman, Buell Jannuzi, Scott Tremaine, 
Andy Gould,  and Peter Goldreich.
Their comments \cite{Video} inspired much of this prefactory: I have stolen from them liberally.  
I am also indebted to Art Poskanzer
and Dick Wolf for sharing their memories of John.   Finally, I especially thank
Neta Bahcall and Jim Peebles
for reading and improving this history.  I am very grateful to Neta for
her encouragement in this endeavor.

\bibliographystyle{arnuke_revised}
\bibliography{bahcall_arxiv}

\end{document}